
\input phyzzx
\overfullrule=0pt
\def\co{cohomology\ }
\def\W{$W_N\ $ }
\def\G{${G\over G}\ $}
\def\GH{${G\over H}\ $}
\def\tGH{twisted\ ${G\over H}\ $}

\def\TKS{twisted Kazama-Suzuki\ }

\def\c#1#2{\chi_{#1}^{#2}}

\def\b#1#2{\beta_{#1}^{#2}}
\def\g#1#2{\gamma_{#1}^{#2}}
\def\r#1#2{\rho_{#1}^{#2}}
\def\jt#1#2{{J^{(tot)}}_{#1}^{#2}}
\def\y{|phys>}
\def\t{\tilde}
 \def\pa{\partial}
\def\bpa{\bar\partial}

\def \jg{J^{(gh)}}
\def \jb{J^{(BRST)}}
\def \dij{\delta_{i,j}}

\def\i{{(ij)}}

\def \vh{\vec{\t {\cal H}}}
\def\cmp#1{{\it Comm. Math. Phys.} {\bf #1}}

\def\pl#1{{\it Phys. Lett.} {\bf #1B}}

\def\prd#1{{\it Phys. Rev.} {\bf D#1}}

\def\np#1{{\it Nucl. Phys.} {\bf B#1}}

\def\jmath#1{{\it J. Math. Phys.} {\bf #1}}
\def\mpl#1{{\it Mod. Phys. Lett.}{\bf A#1}}
\def\jmp#1{{\it J. Mod. Phys.}{\bf A#1}}
\REF\Wym{E. Witten, \cmp {117} (1988) 353, {\bf 118} (1988) 411.}
\REF\EY{ T. Eguchi and N. Yang \jmp  (1990) 1635}
\REF\Li{K. K. Li \np {354} (1991) 711.}
\REF\LVW { W. Lerche, C. Vafa and N. p. Warner \np {324} (1989) 427.}
\REF\Vafa{ C. Vafa, \mpl 6 (1991) 337.}
\REF\us  {O. Aharony,O. Ganor N. Sochen  J. Sonnenschein and S.
Yankielowicz ,`` Physical states in the \G models and two dimensional
gravity", TAUP- 1947-92 April 1992.}
\REF\uss  { O. Aharony, J. Sonnenschein and S.
Yankielowicz ,``  \G models and \W strings" , TAUP- 1977-92 June 1992
(to be published in \pl {} ).}
\REF\HuYu { L. H. Hu and M. Yu, `` On BRST cohomology of SL(2)/SL(2) gauged
WZWN
models" Academia Sinica preprint AS-ITP-92-32.}
\REF\KaSu{ Y. Kazama and H. Suzuki, \np {321} (1989).}
\REF\DG{D. Gepner, \pl {222} (1989) 207; \np{322} (1989) 65. }
\REF\NW{D. Nemeschansky and N. P. Warner ``Topological Matter,
Integrable Models and Fusion Rings"  USC-91/031.}
\REF\NaSu{ T. Nakatsu and Y. Sugawara
`` Topological gauged WZW  models and 2 D gravity" Tokyo preprint  UT-598
``BRST fixed points  and Topological Conformal Symmetry" UT-599.}
\REF\BaRS{K. Bardacki, E. Rabinovici, and B. Serin \np {299} (1988) 151.}
 \REF\GK{ K. Gawedzki and A. Kupianen , \pl {215} (1988) 119, \np
     {320} (1989)649.}
\REF\KS{D. Karabali and H. J. Schnitzer, \np {329} (1990) 649.}
\REF\Wgga{E. Witten, ``On Holomorphic Factorization of WZW and Coset Models"
IASSNS-91-25. }
\REF\Wgg{E. Witten,  `` The N matrix model and the gauged WZW model"
IASSNS-91-26.  }
\REF\Wbh{ E. Witten \prd {44} (1991) 314.}
\REF\Ebh{ T. Eguchi  {\it Mod. Phys. Lett.} {\bf A7} (1992) 85.}
\REF\SY{M. Spigelglas and S. Yankielowicz `` \G Topological Field Theories by
Coseting $G_k$;``Fusion Rules As Amplitudes in \G Theories,''}
\REF\BF  {Bernard   and G, Felder \cmp {}  (1991)  145.}
\REF\GNO{P. Goddard, W. Nahm and D. Olive \pl {160} (1985) 111.  }
\REF\CGHS{ M. Chu, P. Goddard, I. Haliday, D. Olive and A. Schwimmer
\pl {266} (1991) 71.}
\REF\FS { Y. Frishman and J. Sonnenschien \np  {294} (1987) 801 ,
``Bosonization and QCD in two dimensions" TAUP-1981-92.}
\REF\Wak{M. Wakimoto,   \cmp   {104}  (1989) 605.}
\REF\JS{K. Johnson {\it Phys. Lett. }
{\bf 5 } (1963) 253.}
\REF\dvv{R. Dijkgraaf, E. Verlinde, and H. Verlinde,
\np {352} (1991) 59;
``Notes On Topological String THeory And 2D Quantum Gravity,'' Princeton
preprint PUPT-1217 (1990).}
 \REF\BT{ R. Bott L. W. Tu  ``Differential
Forms  in Algebric Topology", Springer-Verlag  NY  1982.}
\REF\BLNW{ M. Bershadsky, W. Lerche, D. Nemeshansky and N. P. Warner
``A BRST operator for  non-critical W strings" CERN-TH  6582/92.}
\REF\BMPN{P. Bowknegt, J. Mc Carthy and  K. Pilch Cern Preprint
TH-6162/91.}
\REF\MS{ M. Speigelglas, \pl {274} (1992) 21.}
\REF\PV{A. Polyakov and P. B. Wigmann \pl {131} (1983)  121.}

\rightline{TAUP-1990 -92}
\date{August 1992}
\titlepage
\vskip 1cm
\title{  On the twisted \GH topological models}
\author {O. Aharony, O. Ganor,  J. Sonnenschein and S. Yankielowicz
\footnote{\dagger}
{Work supported in part by the US-Israel Binational Science
Foundation and the Israel Academy of Sciences.}}
\address{ School of Physics and Astronomy\break
Beverly and Raymond Sackler \break
Faculty of Exact Sciences\break
Tel Aviv University\break
Ramat Aviv, Tel-Aviv, 69978, Israel}
\abstract{
The \tGH models are constructed as twisted supersymmetric gauged WZW models.
We analyze the case of $G=SU(N)$, $H=SU(N_1)\times ...\times
SU(N_n)\times U(1)^r$ with $rank\ G =\  rank\  H$, and discuss possible
generalizations.  We introduce a  non-abelian bosonization of the $(1,0)$
ghost system  in the adjoint of $H$ and in  \GH.
 By computing chiral anomalies in the latter
picture we write  the quantum action as a  decoupled sum  of  ``matter",
gauge and  ghost sectors. The action is also  derived in the unbosonized
version. We invoke a free field parametrization and extract the space of
physical
states by computing
  the  cohomology of  $Q$ , the sum of the  BRST gauge-fixing   charge and the
twisted supersymmetry charge.
For a given $G$ we briefly discuss the relation between the various  \GH models
corresponding to different choices of $H$. The choice $H=G$ corresponds to the
topological \G theory.}
 \endpage

\chapter{ Introduction}
 The $N=2$ super-conformal field theories (SCFT's) have been investigated
intensively during recent years. It is these theories that form
the building blocks in the construction of four dimensional  superstring
theories. Moreover, some of these theories admit Landau-Ginzburg (LG)
description. Upon twisting the $N=2$ theories by the U(1) current in the $N=2$
algebra one obtains a topological theory [\Wym,\EY,\Li ].
 It is still an open question
whether all topological conformal theories are related to $N=2$ super-
conformal theories. The physical  states in the topological
theory are defined  to be in the cohomology of the BRST charge Q.
In the corresponding $N=2$ models, these are precisely the conditions
defining the chiral primary fields.\refmark\LVW Since the LG potential
describes
the chiral fields it also gives, whenever it exists, a solution of the
corresponding topological theory.\refmark{\Vafa} The interplay between the
$N=2$ SCFT and its topological counterpart implies that the topological theory
encodes important information which pertains to the chiral primary
fields. In fact, correlators of these fields can be calculated in the
corresponding topological theory.

 Recently we have investigated the \G topological theories\refmark{\us,\uss}.
We
have shown that there are intriguing similarities between the structure
of \G theories and the structure of 2d gravity theories coupled to $c\leq 1$
matter. As a matter of fact this correspondence goes much further. It
has been  demonstrated\refmark{\us,\HuYu} that upon twisting the \G by a BRST
exact current
$\jt {} 0$, 0ne recovers the 2d gravity coupled to (p,q) minimal models for the
case G=SL(2,R) at level $k={p\over q}-2$. The generalization to G=SL(N,R) at
level $k={p\over q}-N$ gives the $(p,q)$ minimal \W matter coupled to \W
gravity.\refmark\uss
  Among the $N=2$ SCFT's the so called Kazama-Suzuki theories\refmark\KaSu have
attracted much attention. They are based on the coset ${G\over
H}\times SO(dimG/H)$ where \GH is a symmetric space and  the $SO(dim{G\over
H})$
part is associated with the fermionic degrees of freedom which actually reside
in the coset space \GH. We shall refer to the topological version obtained upon
twisting as the \GH topological model. In the present work we will restrict
ourselves to the case G=SU(N) and H a subgroup of G such that $rank(H)=rank(G)$
and investigate the corresponding topological theories. This last requirement
guarantees that \GH is a Kahler manifold which is known to admit $N=2$
supersymmetry. Hence we may take
 $  H=U(1)^{(N-1)}, SU(2)\times U(1)^{(N-2)},......, SU(N-1)\times U(1)$
( Actually we will allow $H=SU(N_1)\times ...\times SU(N_n)\times U(1)^r$
 with $r =N-1-\sum_{I=1}^n
N_I+n$.  ).
Note that the last case  corresponds to the $CP^{(N-1)}$ symmetric
space $ SU(N)/{SU(N-1)\times U(1)}$ which is known to have a LG
description\refmark{\DG,\NW}.
The twist involved in the passage from the $N=2$ theory to the topological
theory turns the coset fermions into  a (1,0) ghost system. If we continue
the series of models  above one step further we will have H=SU(N)
with no coset "fermions". This theory is nothing but the \G topological
theory. Thus, we expect that there exists a strong relationship between
the whole series of topological models depicted above and the \G
theory. In the present work we shall elaborate and analyze this relationship.
Some work in this direction has been done very recently\refmark\NaSu for the
case
$H=U(1)^{(N-1)}$. Related aspects in the symmetric space case $H=SU(N-1)\times
U(1)$ have been investigated in [\NW  ].
 At this stage it is instructive to mention yet another place where
this kind of theories is relevant. This is the black hole case\refmark\Wbh
which corresponds to the CFT coset $SL(2,R)/U(1)$. It has been
shown\refmark\Ebh
that restricting ourselves to the singularity region the theory becomes
essentially a topological theory of the form $U(1)/U(1)$.( At the vicinity
of the singularity the symmetry currents become abelian). In ref. [\Ebh] it
has been shown that yet another description of the singularity is in
terms of the twisted Kazama-Suzuki topological theory $SL(2,R)/U(1)$.
The association of space-time singularities with TFT's of this kind
is another motivation for investigating these theories.

It is a well known fact that \GH coset models can be realized as H-gauged
G-WZW models\refmark{\BaRS,\GK,\KS,\Wgga}. Recently, Witten has demonstrated
that the $N=2$
Kazama-Suzuki \GH models can be formulated as supersymmetric WZW
theories based on the group G where the H subgroup is gauged [\Wgg ].
This will be the starting point of our work.
  The paper is organized as follows: In section 2 we discuss the
non-abelian bosonization of (1,0) ghost systems which appear in the
construction of \GH topological theories. The approach would be to
start with the corresponding (1/2,1/2) fermionic system of the $N=2$
theory, bosonizing it and identifying the U(1) current which twists
the $N=2$
theory into the topological theory. This twist will transform the fermions into
the required (1,0) system. In particular, we will have to deal with fermions
(ghosts) in the adjoint representation of $H$ and in \GH.  Once we have learned
how to bosonize the (1,0) ghost systems we can go ahead and easily  perform
chiral rotations which enable us to decouple  the gauge fields from the ghosts.
This is a crucial  step in the  construction of  the \GH topological quantum
action. This is done,  following essentially the same steps as in the \G
case,\refmark{\GK,\SY}
 in section 3. The outcome is that, as in the \G case,
the action will be composed of three parts. The   matter part is associated
with
G-WZW at level $k$. The  gauge part is associated with $H^{(I)}$
 WZW action
 at level$ -(k+C_G+C_{H^{(I)}})$ for each non abelian gauge factor  $H^{(I)}$
and a    free scalar action with
background charge for each abelian $U(1)$ factor in  $H$ . The third part is a
set  of (1,0) ghost systems. $C_G$ and  $C_{H^{(I)}}$ are the second Casimir
invariants of $G$ and $H^{(I)}$ respectively.
 The ghost part includes the ghosts arising from the $H$-gauge fixing as well
as the   coset ghosts obtained by twisting the $N=2$ fermions.
We repeat the derivation of the quantum action also  in the ``fermionic
language". We use a non-abelian generalization of
Schwinger and Johnson's method to
compute  the chiral anomalies. This results in an action which is
identical to the one derived in the bosonization approach.
  Section 5 is
devoted to the discussion of the algebraic structure of these theories
and its relation to the algebraic structure of the \G theory. We establish
directly that $c^{(tot)}=0$ and identify the zero level Kac-Moody $\jt {} a$
currents associated with the H subgroup. In particular, we have such  a current
per each member of the $G$ Cartan-subalgebra  (recall $rank(H)=rank(G)$).
 As
we will indicate, the algebraic structure of \G is not
precisely the one inherited by twisting the simple $N=2$
algebra.
 For the \GH case the relevant BRST operator will be the sum of
the H-fixing BRST operator and the supersymmetry generator. The
resulting algebraic structure will be a combination of the one
inherited from H/H and the one obtained by twisting the $N=2$
theory. The BRST
cohomology and physical states will be discussed in section 6. The
cohomology of the G/H topological theories will turn out to be
isomorphic to that of the corresponding \G theory. The point is
that on the Fock space the only relations that are crucial for
determining the cohomology are $L_0|phys>=0$
and ${J^{(tot)}}_0^i|phys>=0$, where ${J^{(tot)}}^i$ are the currents in the
G Cartan algebra. Those relations continue to hold in the G/H
topological case. Moreover, the result will continue to hold on
the space of matter irreducible representations since the matter
corresponds to G-WZW of level $k$ in all cases. (Recall that we do not employ
 Felder's procedure\refmark\BF in the H- gauge sector\refmark{\us,\uss}). So,
all
of these theories will have the same number of primary fields and the
partition functions will turn out to be the same.  The obvious
question which arises is to what extent all these models are
equivalent. We will address this question in the last section,
give a summary of our results and make some conjectures. We discuss
the currents needed for the twisting in both the fermionic and bosonic
languages
in an appendix.
\chapter{ Non-abelian  Bosonization  of the \GH ghost systems.}
The ghost systems of the twisted \GH models, as will be  shown explicitly in
the
next section, can be decomposed into two sectors.  The first
one  is an anti-commuting ghost system  of dimensions $(1,0)$ in the adjoint
representation of  $H$. The second sector comprises of   dimension one
anti-ghosts
which correspond to the positive roots of the \GH coset and dimension zero
ghosts associated with  the negative roots. These ghosts are coupled to
non-abelian gauge fields, denoted by $A$, which take their values in the
algebra
of $H$.  The ghost  action, thus,  takes the form:$$ S_{gh}=  -i\int d^2z
Tr_H[\rho\bar D \chi + \bar\rho D \bar\chi] -i\int d^2z \sum_{\alpha\in{ G\over
H}}[\rho^{+\alpha}(\bar D \chi )^{-\alpha}+ \bar\rho^{+\alpha}( D
\bar\chi)^{-\alpha}]  \eqn\mishaghost$$
where $Tr_H$ denotes tracing in the group $H$ and  $D\chi = \pa\chi -
i[A,\chi]$.
The group $H$ is not necessarily  simple. In fact we are interested in the
general case  where the non-abelian   group $H$ can be decomposed into
$H=\prod_I H^{(I)}\times U^r(1)$. Here $ H^{(I)}$ stands for a
non-abelian simple group factor and
 $r=rank\ H-\sum_I rank\ H^{(I)}$. $Tr_H$
therefore means $\sum Tr_{H^{(I)}} $ plus a sum over the $U(1)$ factors.
  The ghosts that correspond to the abelian parts,  are  neutral under
$H$ and therefore they are not coupled to any  gauge fields.   The free ghost
system is obviously invariant under the transformations
  $$\eqalign{\delta \r {}
a &=f^a_{bc}\epsilon^b( z) \r {} c\ \ \ \ \ \delta \c {} a
=f^a_{bc}\epsilon^b( z) \c {} c\cr
\delta \r {} {+\alpha}
&=f^{+\alpha}_{a{+\beta}}\epsilon^a( z) \r {}{+\beta}\ \ \ \  \delta \c {}
{-\alpha} =f^{-\alpha}_{a-\beta}\epsilon^a( z) \c {}
{-\beta},\cr}\eqn\mishad$$ where $f^a_{bc}$ are the structure constants of $H$,
 which are generated by the following currents
 $$J^a= f^a_{bc}\r {} b
\c {} c  + f^a_{+\alpha -\beta}\r {} {+\alpha}
 \c {} {-\beta}. \eqn\mishaJ$$
The indices $a,b,c$ run over the adjoint of $H$ while $\alpha,\beta,\gamma$
denote positive roots of \GH.
 These currents which obey the $H$  Kac-Moody
algebra of level $$k =2C_H + (C_G-C_H)= C_H+C_G\eqn\mishak$$
are coupled to the $H$ gauge field. Obviously there are  anti-holomorphic
currents which  play a similar role.   The energy momentum tensor of the
corresponding free ghost system is  $T(z)= Tr_H[\rho\pa \chi] +
\sum_{\alpha\in{
G\over H}}[\rho^{+\alpha}\pa \chi ^{-\alpha}]$
 and is associated with a Virasoro algebra having a central charge
$$ c= -2d_H-2{(d_G-d_H)\over 2}= -(d_G+d_H).\eqn\mishac$$
Before presenting the  bosonized ghost system it is useful to express the
$(1,0)$ ghost system  in terms of a system of Dirac fermions $(\psi^\dagger,
\psi)$ which transform under $H$ and \GH in the same way as $(\r{} {} , \c {}
{})$. The Kac-Moody level of   the corresponding  fermionic currents  is
identical to that of the ghost system  whereas the Virasoro anomaly is
$c=\half(d_G+d_H)$. Introducing now the twist $T =T_f +\half \pa J_f^\#$, where
the fermion number current  $J^\#_f={\psi^\dagger}^a\psi_a +
{\psi^\dagger}^{+\alpha}\psi_{-\alpha}$, it is easy to check that the Dirac
fermions turn into a
 $(1,0)$ ghost system identical to the one described in eqns.
[\mishaghost-\mishac].

We bosonize first the system of Dirac fermions in the adjoint  representation
of $H$.\refmark\GNO A bosonization is required to produce a bosonic system with
the same
representation of the Kac-Moody and Virasoro algebras as those of the fermionic
system. In addition, one has to check that the sets of primary fields of the
two
formulations are identical. For our purpose of determining the chiral anomaly
it
is enough to fulfill the condition on the Kac-Moody and Virasoro algebras as
well as  to identify  the currents that couple to the gauge fields and the
twisting currents.

 From
here on we will restrict our discussion to the case of $G=SU(N)$ and
$H=SU(N_1)\times ...\times SU(N_n)\times U(1)^r$ with $r =N-1-\sum_{I=1}^n
N_I+n$.  The bosonized    action  takes
now the form $$S^b= \sum_I[ S_{N_I}(l_1^{(I)}, A^{(I)}) +S_{N_I}(l_2^{(I)},
A^{(I)})]+{1\over 2\pi }\int d^2 z  \sum_{s=1}^r\pa\phi^s\bar\pa
\phi^s\eqn\mishaSB$$ where $l_1^{(I)},l_2^{(I)}\in SU(N_I)$.
The Kac-Moody currents associated with $SU(N_I)$ corresponds to the
 sum of the two  WZW  currents  appearing
in \mishaSB. These currents  obviously  have level
$2N_I$.  The total  Virasoro anomaly is (for $A=0$)
 $$c=2\sum_I
{C_{H^{(I)}}d_{H^{(I)}}\over 2C_{H^{(I)}}} + r = \sum_I d_{H^{(I)}}+ r =\sum_I
(N_I^2 -1) + r \eqn\mishacf$$ which is identical to that of the fermions in
the
adjoint of $H$. The operator $Tr[T^a(u_1(z) + u_1^{-1}(z))]$ where $u_1(z)$ is
defined\refmark\CGHS in the appendix, has conformal dimension $\half$.  It is
the bosonic
operator analog to  $\psi_1^a$, where $\psi^a = \psi^a_1 + i \psi^a_2$  and
$\psi_1$ and $\psi_2$ are Majorana fermions.
  Similar relations hold obviously for $\psi_2^a$ and $u_2(z)$.
Thus we can write down the bosonic version of the Dirac field $\psi^a$
 as $Tr[T^a(u_1(z) + u_1^{-1}(z)+ i(u_2(z) + u_2^{-1}(z)))]$.
 To get
to the bosonic version of the $(1,0)$ ghosts one has to twist $T_b\rightarrow
T_b +\half\pa J^\#_b$. The identification of $J^b$ as  a generator for a mixed
symmetry, namely a symmetry that involves the two WZW factors, is given in the
appendix. For each  $SU(N_I)$ group factor  the current
$$J^\#_b=iTr[u_1^{-1}(z)u_2(z) - u_2^{-1}(z)u_1(z)],\eqn\mishJb$$
   is the bosonic counter-part of
$J^\#_f$.
The  holomorphic dimension of the fields which correspond to ${\psi^\dagger}^a$
and $\psi^a$ are shifted from $\half$ to $0$ and  $1$ respectively upon
twisting $T$ with $J^\#_b$
The  Virasoro central charge is shifted by $-3(N_I^2-1)$ thus leading to $c
=-2(N_I^2-1)$  which is equal to that of the ghost system.
 The $U(1)$
currents $J_s=\pa\phi_s$ do a similar job on the abelian parts of $H$.
 The action of the twisted bosonic system, thus, reads
$$ S^b_{(gh)}= S^b-\int d^2 z [ \half\sum_I
J_b^\#\bpa \sigma - \half\sum_s \phi_s \pa\bpa \sigma]\eqn\mishSbgh$$
where the metric is taken to be $h_{z\bar z} =e^\sigma$ so that
$\sqrt{h} R =\pa\bpa \sigma$.

We proceed now to bosonize the coset fermions.
It is easy to describe the \GH
fermions and those in the adjoint of  $H$ by considering
 an $N\times N$ matrix. The latter consists of  $n$ blocks of sizes $N_I\times
N_I$ and $N-\sum_I N_I$  unit blocks   along the diagonal.   The coset fermions
fill the rest of the upper triangle with ${n(n-1)\over 2}$ blocks  of sizes
$N_I\times N_J$,  $r+1-n$ blocks of size $1\times N_I$  for all $N_I$ and
${(r-n)(r+1-n)\over 2}$ unit blocks.  The  fermions  in these off-diagonal
blocks furnish the following
 representations under the non-abelian parts of  $H$: $(1,...,N_I,...,\bar
N_J,...1)$, $(1,...,N_I,...,1)$ and
$(1,...,1)$ respectively, where the numbers inside each
$(...)$ denote the representation under the non-abelian factors .
 The contribution of each block to $c$ is
obviously equal to the dimension of the  corresponding group representation.
The
fermions in the  $N_I\times N_J$ block lead to  $SU(N_I)$ Kac-Moody currents of
level $N_J$, $SU(N_J)$ currents of level
 $N_I$,  and  a fermion number current of level
$N_IN_J$. A similar current structure appears for the one column and unit
blocks.  The bosonized action of such a block is
  $$S^b_{N_IN_J}=S_{N_I}(l_J) +
S_{N_J}(l_I) + {1\over 2\pi  }\int d^2 z \pa\phi_{IJ}\bar\pa\phi_{IJ}
\eqn\mishSni$$
with $l_I\in SU(N_I)$. ( In  the case of one or two abelian factors one can
set
$N_I=1$ or $N_I=1,\ N_J=1$ with $S_1=0$).
Obviously this action results in
the same Kac-Moody
currents as the fermionic ones ($\sqrt{N_IN_J}\pa\phi_{IJ}$ corresponds to
$J^\#$). It is easy to check that  $c=N_IN_J$, as required.
The  bosonic operator which corresponds to a fermion in the $IJ$ block can be
written in terms of  $u_I(z)u_J(z)e^{{i\over \sqrt{N_IN_J}}\phi_{IJ}(z)}$. It
is tempting to write  the bosonic analog of the mass
bilinear  in terms of $l_I$ and $l_J$ as $\psi^{i_I}_{j_J}(z) \bar
\psi_{i_I'}^{j_J'}(\bar z)\simeq (l_I)^{i_I}_{i_I'}(z,\bar z)
(l_J)^{j_J}_{j_J'}(z,\bar z)e^{{i\over \sqrt{N_IN_J}}\phi_{IJ}(z,\bar z)}$.
These operators have the same conformal dimensions and group properties as
those of
free Dirac fermions and the corresponding mass bilinear. However, as was shown
in ref.[\FS] these operators  do not  lead to correlators which are identical
to
those in the fermionic version. It is plausible that  one has to simply use the
bosonization of $\psi$ in terms of $u_I$ and $u_J$. In any case this is not
relevant
 for our present purpose of finding the twisting current which relates only
to the abelian part $e^{{i\over \sqrt{N_IN_J}}\phi_{IJ}(z)}.$
  To transmute a
fermionic block to that of  $(1,0)$ ghosts  we twist  $T$ in the usual
way $T\rightarrow T +\half \pa J^\#$. Changing now the abelian part of
$S^b_{N_IN_J}$ to
$${1\over 2\pi  }\int d^2 z [\pa\phi_{IJ}\bar\pa\phi_{IJ}
+\sqrt{N_IN_J\over 2}\phi_{IJ}R]\eqn\mishsbnn$$
one finds the following contribution to $c$, $c=1-3{N_IN_J}$, so that
altogether
$c=-2N_IN_J$ as that of $(1,0)$  ghosts in the $N_I\times N_J$ block.
The bosonic version  of the ghost system  coupled to $H$ gauge fields  is
easily
acheived  by gauging the bosonic actions constructed  above.

 \chapter{The \GH quantum action using bosonization
of the ghosts. }

Let us start with a derivation of the  quantum action  of the \GH   \TKS model.
 The  classical action  of this model\refmark\Wgg  is that of level $k$ twisted
supersymmetric  $G$-WZW model coupled to gauge fields in the algebra of $H\in
G$.  In other words it is the usual \GH model with an extra  set of $(1,0)$
anti-commuting ghosts where the dimension one fields take their values in the
positive roots of \GH and the dimension zero fields in the negative ones.
The action of the model reads
 $$ \eqalign{S_{(tKS)} &= S_k(g,A,\bar A) + S_{(gh)}^{G\over H}\cr
  S_k(g,A,\bar A)  &= S_{k}(g)
-{k\over 2\pi} \int_{\Sigma}d^2 z Tr_G[g^{-1}\pa g \bar A_{\bar z} +  g\bar \pa
g^{-1} A  - \bar A g^{-1}A  g  + A \bar A  ]\cr
S^{G\over H}_{gh}&=-i\int d^2z \sum_{\alpha\in{ G\over
H}}[\rho^{+\alpha}(\bar D \chi )^{-\alpha}+ \bar\rho^{+\alpha}( D
\bar\chi)^{-\alpha}]\cr}
\eqn\mishwzw$$
where $g\in G$ and $S_k(g)$ is the WZW action at level $k$.
In the case that $H=G$ the model coincides with the \G
model.\refmark\SY\  In fact, we shall follow closely the derivation of the
quantum action of the latter.
In the case that $\Sigma$ is topologically trivial the gauge  fields
can be parametrized as follows $A=ih^{-1}\pa  h ,  \bar  A =i\bar h\bpa
\bar h^{-1}$ where $h(z), \bar h(z) \in H^c$. The WZW part of the  action
then\refmark{\GK,\KS}
takes the form
 $$S_k(g,A) =S_k(hg\bar h)
-S_k(h\bar h) \eqn\mishwzwh$$ The Jacobian  of the change of variables
introduces a   dimension $(1,0)$ system of anticommuting ghosts $\chi$ and
$\rho$ in the adjoint representation of $H$. The WZW  action thus
 becomes
$$S_k(g,A) =S_k(hg\bar h) -S_k(h\bar h) -i\int d^2z
Tr_H[\rho\bar D \chi + \bar\rho D \bar\chi] \eqn\mishwzwh$$ where
$D\chi=\pa\chi -i[A,\chi]$.  One then fixes the gauge by setting $h^*=1$ which
implies $\bar A=0$ and redefining $hg\rightarrow g$. For our case
 where  $G=SU(N)$ and
$H=SU(N_1)\times ...\times SU(N_n)\times U(1)^r$ with $r =N-1-\sum_{I=1}^n
N_I+n$,  the gauge fields $A$ take the form $A= i\sum_{I=1}^n {h^{(I)}}^{(-1)}
\pa h^{(I)} +i\sum_{s=1}^r \pa {\cal H}_s$ and  the \TKS action is given by
 $$ \eqalign{S_{(tKS)} &= S_k(g) - \sum_{I=1}^n S_k(h^{(I)}) -
{k\over 4\pi  }\int d^2 z  \sum_{s=1}^r \pa {\cal H}_s\bpa {\cal H}_s\cr
  &-i\int d^2zTr_H[\rho\bar D \chi + \bar\rho D \bar\chi]
-i\int d^2z \sum_{\alpha\in{ G\over
H}}[\rho^{+\alpha}(\bar D \chi )^{-\alpha}+ \bar\rho^{+\alpha}( D
\bar\chi)^{-\alpha}]\cr}
\eqn\mishwzwHi$$

 The next step taken is to use the results of the previous
section and introduce the bosonized actions of the $H$ and\GH ghost sectors.
The ghost part  in the previous equation is thus replaced by
$$S^b_{(gh)} = {S^b_H}_{(gh)} +\sum_{1\le I<J\le r+1}
{S^b_{N_IN_J}}_{(gh)}\eqn\mishSbb $$ where $S^b_H$ and $S^b_{N_IN_J}$ are given
by eqns. \mishSbgh\ and \mishSni\ (after twisting) respectively. Each of the
non-abelian terms in eqn. \mishSbb\   has the form $$\eqalign{ S_{k'}(\t l,A)
&= S_{k'}(\t l) -{k'\over 2\pi} \int_{\Sigma}d^2 z Tr [ \t l\bar \pa
{\t l}^{-1} h^{-1}\pa h]\cr
 &= S_{k'}(\t lh) -S_{k'}(h)= S_{k'}( l) -S_{k'}(h)\cr}\eqn\mishcSk$$
where we have used the Polyakov-Wiegmann relation and  $l=\t l h$.
The level of $S(h_I)$  in eqn. \mishwzwHi\  is thus shifted
$$-k\rightarrow -k -2 N_I -(N- N_I)=-k-(N+N_I)\eqn\mishshift$$
where the first term $(-2 N_I )$ is due to  ${S^b_H}_{(gh)}$ while the second
one
$[-(N- N_I)]$ comes from $S^b_{G\over H}$.
 Similarly, it is  expected that for  arbitrary groups $G$ and $H_I$ the level
is shifted to $-k-(C_G+C_{H_I})$.
  Now let us examine the coupling to gauge fields of the abelian parts of
$S^b_{(gh)} $.  As was stated above the abelian parts of
 ${S^b_H}_{(gh)}$ are not
coupled to the gauge fields, so we have to discuss only
the coupling of the currents $\bpa\phi_{IJ}$
of each $N_I\times N_J$ block.
These currents couple to a linear combination of the $U(1)$ gauge fields.
Choosing a convenient basis, which fits the  Wakimoto\refmark\Wak free field
parametrization which we will later use for the  non-abelian parts of $H$ (see
eqn. (6.4)) , the latter can be written as $(\vec\alpha_{IJ}\cdot\pa\vh)
\bpa\phi_{IJ}$ where $\vec\alpha_{IJ} = \sum \vec\alpha$ and the sum is over
the
roots which correspond to the block.
 Hence one has to add to $S_{\phi_{IJ}}$ a term of the form
$-\sqrt{2\over N_IN_J}(\vec\alpha_{IJ}\cdot\pa\vh) \bpa\phi_{IJ}$.
The coefficient is determined by the charge of the
fermions in the block with
respect to this abelian gauge field.
 To diagonalize this new action we shift
$\phi_{IJ}\rightarrow   \phi_{IJ} +\sqrt{1\over 2
N_IN_J}\vec\alpha_{IJ}\cdot \vh. $  Under this shift the action is
$$S_{\phi_{IJ}}\rightarrow  S_{\phi_{IJ}}
+{1\over 2 \pi }\int d^2 z[ \half(\vec\alpha_{IJ}\cdot\vh) R -
{1\over 2N_IN_J}(\vec\alpha_{IJ}\cdot\pa\vh)
(\vec\alpha_{IJ}\cdot\bpa\vh) ]\eqn\mishphiij$$
Collecting now all the various terms one finds that  the total
quantum action is  $$ \eqalign{S_k =&S_k(g)\cr
 +&\sum_{I=1}^nS_{-(k +C_G +C_{H^{(I)}}) }(h^{(I)}) +{1\over 2\pi }\int
d^2z[\sum_{s=1}^r \pa {\cal H}_s\bpa {\cal H}_s
 + i\sqrt{2\over
k+C_G}(\vec\rho_G-\vec\rho_H)\cdot \vh R] \cr
&-i\int d^2z Tr_H[\bar\rho \pa \chi + \rho\bar\pa\chi]
-i\int d^2z \sum_{\alpha\in{ G\over
H}}[\rho^{+\alpha}(\bar \pa \chi )^{-\alpha}+ \bar\rho^{+\alpha}( \pa
\bar\chi)^{-\alpha}] \cr}
 \eqn\mishwzwh$$
where we have  normalized the $\vh$ fields to be free  bosons, and
$\vec\rho_G$ and $\vec\rho_H$ are half the sums of the
positive roots of $G$ and $H$ respectively. The action is composed of three
decoupled sectors: the matter sector, the gauge sector and the ghost sector
involving  ghosts in $H$ and \GH. This action  describes a TCFT model as is
shown
in section 5.
 \chapter{ The Quantum action using Schwinger's method}
The derivation of the  quantum action of  the \TKS  model as a sum of
 the matter sector, the gauge sector and the ghost sector which are decoupled
from each other involved a chiral rotation in the ghost sector.
In the previous section a bosonized version of the model was invoked to
perform this transformation which amounts in fact to calculating the chiral
anomaly of the model. Here we repeat this computation in the ``fermionic
language"  for the \GH ghosts. A similar derivation  applies also to the
ghosts in the adjoint of $H$.

Let us define
$$ e^{iI(h,\bar  h)} = \int D\rho D\bar\rho D\chi D\bar\chi
e^{ {i\over 2\pi}\int
d^2z \sum_{\alpha\in{ G\over H}}[\rho^{+\alpha}(\bar D \chi )^{-\alpha}+
\bar\rho^{+\alpha}( D \bar\chi)^{-\alpha}]}\eqn\misheh$$
where again we parametrize $A=ih^{-1}\pa  h ,  \bar  A =i\bar h\bpa
\bar h^{-1}$ where $h(z)\in H^c$. Under the  infinitesimal chiral
transformations
$\delta h =\epsilon h,\ \delta \bar  h = \bar h\bar\epsilon$  we have
$$\eqalign{\delta e^{iI(h, \bar h)}\over e^{iI(h, \bar h)} &= -< {1\over
2\pi}\int d^2z \sum_{\alpha\in{ G\over H}}[\rho^{+\alpha} [\delta \bar A,
\chi]^{-\alpha}+ \bar\rho^{+\alpha} [\delta A, \bar\chi]^{-\alpha}]>_{h\bar
h}\cr
&= -{1\over 2\pi}\int d^2z \sum_{\alpha\in{ G\over H}}f_{a\alpha,-\beta}
[<\rho^\alpha \chi^{-\beta}>_{h,\bar h}\delta\bar A^a +   <\bar\rho^\alpha
\bar \chi^{-\beta}>_{h,\bar h}\delta  A^a\cr}\eqn\mishdel$$
 From here on we consider a
chiral transformation only  of $\bar h$, namely $\epsilon=0,\bar\epsilon \ne
0$.  This last relation is meaningful only provided that we regularize in a
gauge invariant way the propagator $<\r {} \alpha (z,\bar z)\c {} {-\beta}(z,
\bar z)>$. This can be acheived by generalizing a method proposed originally by
Schwinger and Johnson\refmark\JS for the abelian case namely:
 $$<\r {} \alpha (z,\bar z)\c {} {-\beta}(z, \bar
z)>_{h \bar h}= lim_{z'\rightarrow  z} {\cal G}^{\alpha ,-\gamma}[P
e^{i\int_{z'}^z [A(z",\bar z")d \bar z" + \bar A(z",\bar z")d
z"]}]^{-\beta}_{-\gamma},\eqn\mishG$$
 where the Green's function ${\cal G}^{\alpha ,-\beta}$ is a solution of the
equation
$$ \bar D_{z'} {\cal G}^{\alpha ,-\beta}= \bpa_{z'} {\cal G}^{\alpha ,-\beta}
-if_{\ a\gamma}^{\alpha} \bar A^a({z'})  {\cal G}^{\gamma ,-\beta} = \pi
\delta^{\alpha
\beta}\delta^{(2)}({z'}-z).\eqn\misheG$$
 Some clarifications are in order: (i)
${z'},z$ are on the plane ${\cal C}-\{0\}$ and we use the equal radius limit
$|{z'}|=|z|$ (which may be viewed as an equal time limit on a cylinder). (ii)
The
path ordered integral is not uniquely defined since the path is not defined.
However, the ambiguity is of order $O(|{z'}-z|^2)$ and, therefore, is
negligible
since ${\cal G}\sim {1\over {z'}-z}$. (iii) For any $b\in H$
$[b]_\alpha^\beta \equiv Tr[T^{-\gamma} b^{-1} T^\beta b]g_{\alpha,-\gamma}$.
It is straightforward to prove that eqn. \mishG\ is indeed gauge invariant. Now
the solution of  equation \misheG\  is given by
$$ {\cal G}^{\alpha ,-\beta}(z',z)= [\bar h(z')]^{\alpha}_{\alpha'}{{\cal
G}^0}^{\alpha', -\beta'}(z',z)[\bar h(z)]^{-\beta}_{-\beta'}\eqn\mishsG$$
 where ${{\cal G}^0}^{\alpha, -\beta}=
{g^{\alpha,-\beta}\over {z'}-z} $ is the free propagator.
We take now the limit in eqn. \misheG\ and find
 $$<\r {} \alpha (z,\bar z)\c {} {-\beta}(z, \bar
z)>_{h \bar h}= <\r {} \alpha (z,\bar z)\c {} {-\beta}(z, \bar
z)>^0  + f^{a\alpha,-\beta} [(\pa
\bar h\bar h^{-1})_a - ( h^{-1}\pa h)_a)]\eqn\mishrc$$ Inserting this result
into
eqn. \mishdel\ we get
$$\eqalign{\delta I(h,\bar h) &=
 {i\over 2\pi}\int d^2z \sum_{\alpha\in{ G\over
H}}f_{a\alpha,-\beta}<\r {} \alpha \c {} {-\beta}>_{h,\bar h}(\bar  h \bpa
\bar \epsilon \bar h^{-1})^a  \cr
&= {i\over 2\pi}\int d^2z \sum_{\alpha\in{ G\over
H}}f_{a\alpha,-\beta}[<\r {} \alpha \c {} {-\beta}>^0
+ if^{b\alpha,-\beta}( (\pa\bar h \bar h^{-1})_b+ ( h^{-1}\pa h)_b)](\bar
h\bpa \}
bar \epsilon \bar h^{-1})^a  \cr}\eqn\mishdee$$
Using the non-conservation of the fermionic current due to  the gravitational
anomaly, namely $\bpa (\r {} \alpha \c {} {-\beta}) = \half \delta^{ \alpha,
\beta} R $ one finds that
$$ I(h,\bar h) = S_{(C_G-C_H)} (h\bar h)  +{1\over 2\pi }\int i\sqrt{2\over
k+C_G}(\vec\rho_G-\vec\rho_H)\cdot \vh R \eqn\mishII$$
where we have used  $\sum_{\alpha,-\beta \in {G\over H}}f^i_{\alpha,-\beta}
\delta^{\alpha,\beta}= \rho^i_G-\rho^i_H$ and $
f^i_{\alpha,-\beta}f_j^{\alpha,-\beta} = (C_G\delta_j^i-{C_H}_j^i)$ and where
${C_H}^i_j=C_{H_I}\delta^i_j $ if both $i$ and $j$ are in $H_I$ and otherwise
zero.
Thus the fermionic derivation of the chiral transformation of the quantum
action is identical to that derived using the bosonization method.
 \chapter{ The algebraic structure of the \TKS model}
The next step in analyzing the \tGH models is the derivation of their algebraic
structure.  Twisted $N=2$ super conformal models,  as well as  the \G models,
obey the TCFT algebra.\refmark\dvv  As one would expect the
\TKS  models share the same algebra.
We start with the Kac-Moody algebra associated with  the group $H$. We define
the currents ${\jt {} {a}}_I$ for each  non-abelian group factor  $H^{(I)}$,
and
for each $U(1)$ group as
 $${\jt {} {a}}_I   =J^a +I^a  +if ^a_{bc}\r {} b \c {} c
+i\sum_{\alpha\beta\in {G\over H}} f^a_{\alpha,-\beta}\rho^\alpha\chi^{-\beta}.
\eqn\mishbJ$$  where $J^a,\ I^a$  are the contributions of the $g$ and $h$
sectors respectively.  The contribution of the   ghost currents
consists of both
the $H$ and \GH parts to be denoted by $J^a_H$ and $J^a_{G\over H}$
respectively.
  The level of these currents vanishes
 $$ k_I^{(tot)} = k
-(k+C_G+C_{H_I} ) + 2C_{H_I} +(C_G-C_{H_I}) =0. \eqn\mishbk$$ For the
abelian case there is a similar expression now with  $C_{H_I}=0$.

The energy momentum tensor $T$ can be decomposed, in a way which will be found
later to be natural,  into $T=T^H+T^{G\over H}$ as follows
$$ \eqalign{ T(z) &= {1\over2( k+c_G )} g_{\t a\t b}:J^{\t a} J^{\t b}: -
{1\over 2(k+c_G )}g_{ab}:I^a I^b: + g_{ab}\r  {} a \pa \c   {} b
\cr &-{\sqrt{2}\over k+C_G}(\vec\rho_G-\vec\rho_H)\pa\vec I
+ \sum_{\alpha\in{ G\over
H}}\rho^{+\alpha}( \pa \chi )^{-\alpha}\cr
T^H(z) &= {1\over2( k+c_G )} g_{ab}:(J^a +J^a_{G\over H})(J^b+J^b_{G\over
H}):-{1\over 2(k+c_G )}g_{ab}:I^a I^b:
\cr&-{\sqrt{2}\over k+C_G}(\vec\rho_G-\vec\rho_H)\cdot\pa(\vec J +\vec I +\vec
J_{G\over H})
 + g_{ab}\r  {} a \pa \c   {} b\cr
T^{G\over H}(z) &= {1\over 2(k+c_G )}g_{\t a\t b}:J^{\t a} J^{\t b}: -{1\over2(
k+c_G )} g_{ab}:(J^a +J^a_{G\over H})(J^b+J^b_{G\over H}): \cr
&+{\sqrt{2}\over
k+C_G}(\vec\rho_G-\vec\rho_H)\cdot\pa(\vec J  +\vec J_{G\over H})
+\sum_{\alpha\in{ G\over H}}\rho^{+\alpha}( \pa \chi )^{-\alpha}
 \cr}\eqn\mishbT$$
where $\t a$ and $\t b$ go over the adjoint of  $G$.
$\vec J$, $\vec I$ and $\vec J_{G\over H}$ are the Cartan-subalgebra currents
given in the basis in which $[J^i_n, J^j_m] = k n \delta ^{ij}\delta_{m+n}$.
The total Virasoro central charge vanishes since

$$ c= {kd_G\over k+C_G} +\sum_{I=1}^n {(k+C_G+C_{H^I})d_{H^I}\over k+C_G} +r
-2d_H -(d_G-d_H) +6[\sqrt{2\over k+C_G}(\vec\rho_G-\vec\rho_H)]^2 =
0\eqn\mishcz$$ where we have  used, assuming simply laced groups,
 the relations  $12\rho_G^2 =d_GC_G$, and
$\vec\rho_H\cdot (\vec\rho_G-\vec\rho_H)=0$

In addition to the commuting holomorphic ( and anti-holomorphic) symmetry
generators there are also anti-commuting ones. Upon gauge fixing, the gauge
invariance is transformed into a BRST symmetry generated by a dimension one
current $\jb$. This current has a dimension two partner $G$. These two
anti-commuting currents are given by
$$ \eqalign{ \jb=&g_{ab}\c  {} a[J^b + I^b + J_{G\over H}^b + \half {\jg}^b_H]
=g_{ab}\c  {} a[J^b + I^b + {i\over 2}f^b_{cd}\r {} c \c {} d  + {i}
f^{b}_{\gamma,-\beta}\rho^\gamma\chi^{-\beta})\cr
G^H=&{g_{ab}\over 2(k +c_G)}\r  {} a [J^b - I^b +
if^{b}_{\gamma,-\beta}\rho^\gamma\chi^{-\beta}]
-{\sqrt{2}\over
k+C_G}(\vec\rho_G-\vec\rho_H)\cdot\pa\vec\rho
  \cr}\eqn\mishbGJ$$ It is
straightforward to realize that $T^H(z)$ is BRST exact
$$T^H(z)=\{Q^{(BRST)},G^H(z)\} \eqn\mishQGT$$ where $Q^{(BRST)}=\oint dz
J^{(BRST)}$. As was shown in ref.[\Wgg] the addition of the coset ghosts
turned the model into a twisted $N=2$ model. The twisted $N=2$ algebra is
generated
by $Q^{G\over H}$ and $G^{G\over H}$ given by
their $N=2$ counterparts\refmark\KaSu  with the fermions replaced with ghosts.
$$\eqalign{ Q^{G\over H}=
\sum_{\alpha\beta\gamma\in {G\over H}}\c {} {-\alpha}(J^\alpha + {i\over 2}
f^{\alpha}_{\gamma,-\beta}\rho^\gamma\chi^{-\beta})\cr
G^{G\over H}={1\over k+C_G}\sum_{\alpha\beta\gamma\in {G\over H}}\r {}
{\alpha}(J^{-\alpha} + {i\over 2}
f^{-\alpha}_{\gamma-\beta}\rho^\gamma\chi^{-\beta}).\cr }\eqn\mishQG$$
$T^{G\over H}$ defined above is exact with respect to $ Q^{G\over H}$
$$T^{G\over H}=\{Q^{G\over H},G^{G\over H}\}. \eqn\mishQGH$$
The various  $Q's$ and $G's$ obey the following anti-commutation relations:
$$ \eqalign{\{Q^{G\over H},Q^{(BRST)}\}&=
\{Q^{G\over H},G^H\}=\{Q^{(BRST)},G^{G\over H}\}= 0\cr
\{Q^{(BRST)},Q^{(BRST)}\}&= \{Q^{G\over H},Q^{G\over H}\}=\{G^{G\over
H},G^{G\over H}\}= 0\cr
\{G^{H},G^H\}&={1\over 4(k+C_G)^2} f_{abc}\r {} a  \r {} b \jt {} c
\cr}\eqn\mishGGQQ$$ The fact that $G^H$ is not nilpotent is shared by several
other TCFT's, in particualr the \G models.
  Defining now the combined generators
$$ Q= Q^{(BRST)} +Q^{G\over H}\qquad  G= G^H + G^{G\over H} \eqn\mishQG$$
we find the following relations which are common to all
 TCFT algebras\refmark\dvv
 $$\eqalign{ T(z) =&\{ Q, G(z)\} \cr  \jb =&\{ Q ,
j^\#(z)\}\cr
\jt {} a =&\{ Q ,\r {} a \}\cr}
\eqn\mishalgebra$$
where $\jt {} a$ denotes a current in the  algebra of $H$ and
$j^\#(z)= g_{ab}\r {} a \c {} b +\sum_{\alpha\in{ G\over
H}}\rho^{+\alpha} \chi^{-\alpha}$. Notice that $G$ is
not nilpotent.  Following eqn.\mishGGQQ\  it is clear that the algebra given in
 eqn.\mishalgebra\ is not closed. A complete analysis of the algebraic
structure
is under current investigation.

\chapter { The $Q$ cohomology  of the \tGH model}
Next we  proceed to extract  the space of  physical states of the model.
We take as our definition of a physical state a state
 in  the  \co of  $ Q= Q^{(BRST)} +Q^{G\over H}$ ,namely, $|phys>\in
H^*(Q)$. In  the case that
the spectral sequence\refmark\BT of the double complex of $Q^{(BRST)}$
and   $Q^{G\over H}$ degenerates at the $E_2$ term, this is the same as taking
one cohomology and then the other. It is not clear to us whether this always
happens in our case.
   Expanding the various currents in modes, $Q$ takes the form
$$\eqalign{ Q&=\sum_{n, m=-\infty}^\infty [ g_{ab}\c n a( J_{-n}^b+I_{-n}^b)
-\half f_{abc}  :\c{-n} a\c{-m}  b  \r{n+m} c: ]\cr
+& \sum_{\alpha\beta\gamma \in {G\over H}}\sum_{n, m=-\infty}^\infty
[g_{ab}f^b_{\alpha,-\beta}
 :\c{-n} a\r{n+m} \alpha\c{-m}  {-\beta}
:  + \c {n}  {-\alpha}( J_{-n}^\alpha+\half  f^\alpha_{\beta,-\gamma}
 :\r{-m} \beta \c {m-n} {-\gamma} ):
]\cr} \eqn\mishQ$$
The extraction of the physical states from here on follows the same lines as in
 the \G models described in details in refs. [\us,\uss].  Therefore, here we
only
  briefly summarize the procedure.
The physical states obey
$$L_0\y =  0 \qquad \jt 0 i \y =  0 \  \ \ \  (i=1,..., rank\  H)\eqn\mishL.$$
  The  generalized BRST charge  is decomposed into
$$\eqalign{Q  &= \c 0 i \jt 0 i  + M^i \r 0  i +\hat Q \cr M^i &= -\half
f^i_{bc}\sum_{n}  :\c{-n} b\c{n}c : \cr} \eqn\mishhatQ$$
where the  sum over $i$   is over the Cartan subalgebra,
so that on the sub-space  of states annihilated by   $\r
0 i$,  $Q=\hat Q$. The cohomology on this subspace is called the relative
cohomology. Let us start to compute this cohomology.
   The states of  $H^*(\hat Q)$ are built  on a highest weight vacuum  $|J,I>$
defined in ref. [\us,\uss] by applying  creation operators which correspond to
a
free field parametrization of the $J$ and $I$ sectors.  (The root
denoted in ref. [\uss] by $(ij)$ correspond to $\alpha_i+...+\alpha_j$). Notice
that the bosonization of the $I$ sector is opposite to  the one used in
the $J$ sector. In both cases there are bosons which correspond to the Cartan
sub-algebra denoted by ${\cal H}_i$ and $\t{\cal H}_i$ for the $J$ and $I$
sectors respectively. In addition there is a pair of dimension  $(1,0)$ ghost
for
each root. The dimension one fields denoted by $\beta$ relate to the positive
roots in the $J$ sector, while their analogs $\t\beta$ are associated with the
negative roots in the $I$ sector.
  For instance in this parametrization the components of the $I$
currents which are associated with the Cartan sub-algebra of the $H_I=SU(N_I)$
group factor
 take the form\refmark\uss ( in a basis which satisfies
 $[I^i_n, I^j_m] = -(k+ N+N_I) n  g^{ij}\delta_{m+n}$
$$ \eqalign{
I_n^i =& -\sum_{j=1}^{N_I-1} g_{H_I}^{ij} [i\sqrt{k+N}\vec\alpha_j\cdot
\vec{\t{\cal H}}_n + 2\sum_m :\t\g m {(jj)} \t\b {n-m} {(jj)}:\cr
-&\sum_m\sum_{k=1}^{j-1}(:\t\g m {(k,j-1)}\t\b {n-m} {(k,j-1)}:-:\t\g m {(kj)}
\t\b {n-m} {(kj)}:)\cr
 -&\sum_m\sum_{k=j+1}^{N_I-1}(:\t\g
m {(j+1,k)}\t\b {n-m} {(j+1,k)}:-:\t\g m {(jk)} \t\b {n-m}
{(jk)}:)].\cr}\eqn\mishJij$$ Here $\t{\cal H}_i, \
i=1,...,N_I-1$  are scalar fields
with $\t{\cal H}_i(z)\t{\cal
H}_j(\omega) =-\dij log (z-\omega)$
 and  $\t\b {} \i , \t\g {} \i $ $(i\leq j)$ are the commuting $(1,0)$
ghost  systems
obeying $\t\g {} \i (z)\t\b {} {(kl)} (\omega) =\delta^{ik}\delta^{jl} {1\over
(z-\omega)}$.
 Defining degrees and using a spectral sequence decomposition
for $\hat Q$ , it turns out\refmark\uss that the required \co is
isomorphic to that  of the zero degree
 component of $Q$.  The contribution of the excitations of various fields to
$L_0$ are exact under the latter operator and hence there are no excitations in
$H^*(Q)$.  Finally one finds\refmark\uss
 that the  relative cohomology  contains only a
single state given by
 $$ H^{rel} (Q) = \{\prod_{\alpha\in H,\alpha>0}  \c 0
\alpha|\vec J, \vec I>
 ;\   \ \ \vec J + \vec I + 2 \vec \rho_H=0\} .
\eqn\mishrelco$$
Since  the $J$ sector has a  background charge of ${i\over  \sqrt{k+C_G}}\vec
\rho_G\cdot   \pa^2\vec\phi$ while in  the $I$ sector has a background charge
of
${i\over  \sqrt{k+C_G}}
(2\vec\rho_H-\vec\rho_G)\cdot   \pa^2\vh$, we find that the weight of this
state is
$L_0= {1\over   k+c_G}[\vec J\cdot (\vec J +2\vec \rho_G)-
\vec I \cdot (\vec I +2(2\vec \rho_H-\vec \rho_G))]=0$ for
 $\vec J + \vec I + 2 \vec
\rho_H=0$. This state  corresponds to the ``tachyon" state of the $W_N$ models
based on $H=G=SU(N)$.\refmark\BLNW
 The absolute cohomology (without the restriction
$\r 0 i=0$ ) is $$H^{abs} (Q) \simeq H^{rel} (Q)
\oplus\sum_{\{k_1,...,k_l\}}\c 0 {k_1}... \c 0 {k_l} H^{rel} (Q)
\eqn\mishabsco$$ where the sum  is over $\{k_1,...,k_{l}\}$ which are  all
possible subsets of the set  $1,...,N-1$. Thus, each state in the relative
cohomology gives rise to $2^{N-1}$  states in the absolute cohomology.
In the \G models the physical states were deduced after  translating  the
  cohomology
on the Fock space into the space of  irreducible representations of the
Kac-Moody algebra associated with the  $G$-WZW
matter sector.\refmark{\us\uss}
Following the same steps in the present case one finds for each maximal weight
$J$ of $G_k$ a $rank\  G$ dimensional vector of states. This implies also that
there is an $N-2$ dimensional lattice of states   for each ghost number and
$J$.
The latter situation  follows
from the $N-1$ dimensional lattice of Fock spaces which are derived by Weyl
reflections as well as shifts by  linear combinations of  roots.\refmark\BMPN
A correspondence between the field content ( up to ``topological sectors") and
 partition functions  of the $(p,q)$  $W_N$ strings and \G models for
$A^{(1)}_{N-1}$  at level
 $k={p\over q}-N$
(the case of $N=2$ corresponds to the
 minimal models coupled to gravity)
was derived\refmark{\us,\HuYu}
by  further  twisting the  models according to
$$T(z)\rightarrow \t T(z) = T(z) +\sum_{i=1}^{N-1}\pa\jt {} i(z).\eqn\mishetT$$
It is easy to realize that the torus partition sum in the
present case is equal to the one of the  $A^{(1)}_{N-1}$ and thus shares the
same
relation with the $W_N$ string models.\refmark\uss

\chapter{ Summary and discussion}
In this paper we have analyzed  the \tGH models.
 We focused  on the case of $G=SU(N)$, $H=SU(N_1)\times ...\times
SU(N_n)\times U(1)^r$ with $rank\ G =\  rank\  H$, and discussed possible
generalizations to other groups.   The quantum action  was written as a
decoupled sum  of  ``matter",  gauge and  ghost sectors.  This was achieved by
computing chiral anomalies both  by introducing a  non-abelian bosonization of
the $(1,0)$
ghost system  in the adjoint of $H$ and in  \GH, as well as in the unbosonized
version.  The algebraic structure of the models was presented emphasizing its
relations with the TCFT algebra.\refmark\dvv
 We invoke a free field parametrization and extract the space of physical
states by computing
  the  cohomology of  $Q$ , the sum of the  BRST gauge-fixing   charge and the
twisted supersymmetry charge.

The exact relations between models based on the same $G$ but different $H$ can
be understood only after computing all possible correlators for the various
models. Even though this stage of understanding the models is still ahead of
us,
we can make the following remarks.

(i) The field content of the various models is very similar.
Two such \tGH models  differ in quartets
of fields $\c {} \alpha, \r {} {-\alpha}, \t\b {} {-\alpha},
\t\g {} \alpha$, corresponding to positive roots $\alpha$ which are
conained  in one  $H$
subgroup and not in the other. Those
 quartets contribute nothing to the
Virasoro anomaly .

(ii) The cohomologies of the various cosets are very similar. When an inverted
bosonization was used for the gauge sector
 ( as was done in section 6), the Fock-space cohomology
was found to  consist of one state for every eigen-value  $\vec J$ of  the
Cartan subalgebra. The $\vec I$  eigenvalue was connected to
that of  $\vec J$ via $\vec I+\vec J +2\rho_H=0$ and the ghost number of the
state was equal to the   number of positive roots of $H$.  It is thus clear
that there is a
simple isomorphism between the cohomologies of the different coset models
which preserves the $\vec J$ and $L_0$ eigenvalues of the states.

(iii) A similar relation holds  when  we work out
the cohomology  of the  physical states
in the space of
irreducible representation of the Kac-Moody algebra of the matter
sector.  Since the matter sector is identical for all the various  choices
of $H$ we find, after applying the method of ref. [\BF],  that the states of
the different models are connected by a constant shift of $I$ and of the
ghost number.

(iv) As an immediate outcome of the previous points, it is  clear that the
torus partition sum  is the same for all subgroups $H$.

(v)  Non-trivial correlation functions on the sphere  are those that obey the
condition that the sum of the ghost number of the operators is equal to the
ghost number anomaly. The latter is proportional to the rank of the group and
thus is the same
for the various  cosets.
(vi) The LG approach, whenever it exists, gives a description of chiral
primary fields of the $N=2$ theory and, therefore, of the primary fields  of
the
corresponding twisted topological theory.  Thus the LG theory gives a solution
of the corresponding topological theory.  We conjecture that all the \tGH
models, including the \G model, are connected by smooth perturbations. In the
LG approach
this means that the leading power in all the corresponding potentials  is the
same. Thus,  the different \tGH models would  correspond  to different points
in the space of parameters associated with the $LG$ potentials,
  at which the model is both conformal and topological.
This conjecture is supported by some specific examples. As an illustration take
the case of $A_k$ minimal $N=2$ models based on ${SU(2)_k\over U(1)}$ coset.
The corresponding TCFT\refmark{\EY,\Li,\dvv}
 are described by the potential $W(x) = {x^{k+2}\over k+2}$ and the primary
fields are $\phi_i\sim x^i$ with $i=0,..., k$
They satisfy the chiral ring realation $\phi_i\cdot \phi_j = \phi_{i+j} \
mod(k)$
associated with the ring of polynomials $R={C[x]\over dW}$.
 One can perturb this
potential by turning on a set of couplings $\{ t_n\}$ to the operators
$\phi_n$.
There is a well defined\refmark\dvv iterative way to construct the perturbed LG
potential. By this procedure one obtains  a multi-parameter family o
fpotentials
$W(x,\{t_n\} )$ which in this example takes the form $ W(x,\{t_n\} )=
{x^{k+2}\over k+2}- \sum_{i=0}^k g_i(t) x^i$, with well defined coefficients
$g_i(t)$ ( $g_i(t)= t_i +...$). The corresponding primary fields are given by
$\phi_j(x,\{t_n\} )= -{\pa W(x,\{t_n\} )\over \pa t_j}$ and they satisfy the
ring structure $\phi_i(x,\{t_n\} )\cdot \phi_j(x,\{t_n\} ) = \sum_k c_{ij}^k
\phi_{k}(x,\{t_n\} )$.
At the origin of the $\{t_n\}$ parameter space we have the twisted
${SU(2)_k\over U(1)}$  theory\refmark{\EY,\Li}.
 At the point $t_i=\delta_{i,k}$
we obtain the \G model for $SU(2)$ level $k$.\refmark{\MS}
Note that these two points are connected along the line $t_i=t\delta_{i,k},
0\leq t\leq 1$. At the point $t=1$ , $\phi_i(x,t=1)= P_i(x)$ where the $P_i's$
are the modified Chebyshev polynomials. These polynomials subject
to  the constraint
$W'(x, t=1) \equiv P_{k+1}(x) =0$ are known to satisfy the fusion ring
structure\refmark{\MS,\DG} i.e. $C_{ij}^k (t=1) = N_{ijk}$. Other specific
examples that support our conjecture can be worked out given the LG potentials
given  in ref. [\NW]. It would be
very interesting to prove our conjecture, identify and provide chracterization
to all the points in the LG  parameter space
 that correspond to TCFT's. Note
that upon twisting the latter one can get the LG description of the
correponding $N=2$ theories which for most of the cases under consideration are
not known yet.
\ack { We would like to thank T. Eguchi for  a stimulating
discussion.
 One of us J.S would like
to thank A. Schwimmer for a useful conversation.}
\refout
 \appendix{} {Mixed symmetries in non-abelian bosonizations}
Let us start with the simple case of abelian bosonization of  two
Dirac fermions. The fermionic action is
$$S=\int d^2z \sum_{i=1}^2[\psi_i^\dagger\bpa\psi_i +  \bar\psi_i^\dagger\pa
\bar\psi_i]\eqn\misha$$
Each sector is obviously invariant under holomorphic ( and anti-holomorphic)
transformations generated by $J^f_i=\psi_i^\dagger\psi_i$ and
$T^f_i=\half[\psi_i^\dagger\pa\psi_i-\pa\psi_i^\dagger\psi_i]$. In addition the
current  $$ J^f=\psi_1^\dagger\psi_2 -\psi_2^\dagger\psi_1 \eqn\mishb$$ is
also a holomorphic current.  Its dimension is (1,0) and $\bpa J^f=0$ since the
fermions $\psi, \psi^\dagger$  are all holomorphic by the equation of motion.
Under $J^f$ one has the transformations:
$$\delta\psi_1=-\epsilon(z)\psi_2\qquad \delta\psi_2=\epsilon(z)\psi_1
\eqn\mishc$$ which obviously leave the action invariant.

The corresponding bosonic system has the action
 $$S=\int d^2z \sum_{i=1}^2 \pa\phi_i\bpa\phi_i \eqn\mishd$$
Again the symmetries of each sector are $J^b_i=\pa\phi_i$  and $T^b_i
=\pa\phi_i\pa\phi_i$  and in addition there are some mixed symmetries. We want
to identify what is the symmetry which corresponds to $J^f$. The
transformations $\delta \phi_1=-\epsilon \phi_2 \qquad\delta \phi_2=-\epsilon
\phi_1$ leave the action invariant but the corresponding current
$$ J^b= \phi_1\pa\phi_2-\phi_2\pa\phi_1 , \qquad
\bpa J^b=\bpa\phi_1\pa\phi_2-\bpa\phi_2\pa\phi_1\ne 0 \eqn\mishe$$  is not
holomorphically conserved.
 The corresponding vector current is conserved namely:
$\bpa J^b + \pa \bar J^b= 0$  where $\bar J^b =
\phi_1\bpa\phi_2-\phi_2\bpa\phi_1$.
The difference between the chiral and vector  conservations manifests itself in
the nature of $\epsilon$ the parameter of transformation. The action is
invariant under global transformations only and not with $\epsilon(z)$ or
$\bar \epsilon(\bar z)$. Notice that unlike the usual currents $J^b_i$ where
the
conservation of $J^V_\mu$ implies the conservation of
$J^A_\mu=\epsilon_{\mu\nu} {J^V}^\nu$, here this relation does not hold.
 This is
of course equivalent to the non-invariance of the chiral current.
It is thus obvious that this current cannot correspond to $J^f$.
The bosonic version of $J^f$ can be obtained by bosonizing separately $\psi_1$
and $\psi_2$.  The result is given by
$$J^b=:e^{-i\phi_1}::e^{i\phi_2}:+:e^{-i\phi_2}::e^{i\phi_1}:= 2
:cos(\phi_1-\phi_2):.\eqn\mishf$$ Notice that here $\phi=\phi(z)$.
(This can   be
written in Mandelstam formulation). Therefore, by construction $\bpa J=0$.
Using the transformations
 $$\delta \pa\phi_1=-\epsilon(z) :sin(\phi_2-\phi_1): \qquad
\delta \pa\phi_2=\epsilon(z) :sin(\phi_2-\phi_1): \eqn\mishg$$
it is easy to verify that the action is indeed invariant:
$\delta S^b=\int d^2z \bpa  :sin(\phi_2-\phi_1):=0$. In addition the dimension
of
$J^b$  is $(1,0)$. So we have identified the bosonized version of $J^f$.

Let us now pass to the non-abelian case.
As discussed in section 2  the bosonized theory of Dirac fermions in the
adjoint representation  is the sum of two WZW models at level $N$ for
$H=SU(N)$.
 Following   the discussion  above, in addition to the symmetries of
each sector there are also some mixed symmetries. The current relevant to the
twisting is
$$ J_\#^f= Tr_H[\psi_1{\psi}_2]=\psi_1^a{\psi_a}_2  \eqn\mishb$$
For each index $a$ the transformations are those given in eqn.\mishc. The
Kac-Moody level of this current is $N^2-1$.
In analogy to the abelian bosonization we construct the operator
$$J^\#_b=iTr[u_1^{-1}(z)u_2(z) - u_2^{-1}(z)u_1(z)],\eqn\mishJbbb$$
The chiral  fields  $u(z),\bar u(\bar z)$ are the non-abelian analogs of
$:e^{i\phi(z)}:$ and $:e^{i\bar \phi(\bar z)}:$. Classically, the
solution of the equation of motion can be written as
$g(z,\bar z)=u(z)\bar u (\bar z)$ just like
$\phi(z,\bar z ) = \phi(z) +\t\phi(\bar z)$
for the abelian case. The quantization of these chiral
group elements, including a careful treatment of the ambiguities involved in
their definitions is discussed in ref. [\CGHS].
For $u_1,u_2$ in the adjoint representation of $SU(N)$ at level $N$, the
dimension of $u(z)$ is $\Delta = {C_G \over k+N} = {N\over N+N}=\half$. Thus,
 $J_H^b$
has dimension $(1,0)$( by construction $\bar\pa  J_H^b =0$).
 Notice that
the Kac-Moody algebra of $ J_H^b$ is of level $N^2-1$ so it matches that of
 $J_H^f$.

Twisting the enegy momentum tensor with $\pa J_H^b$ produces a shift of
$-3(N^2-1)$ as requested. So this twisted $T$ leads to  the $(1,0)$ system in
the
adjoint representation. To introduce an action that corresponds to this
energy-momentum tensor, it seems that one has to write some non-local term.
 If $ J_H^b=\pa\phi$ then we can  add a term of the form $\phi R$,
alternatively if $R=\pa\bpa \sigma$, then one can add the term $
J_H^b\bpa\sigma$.

\end